\documentclass{article}

\usepackage[preprint]{style_2024}

\usepackage[utf8]{inputenc} 
\usepackage[T1]{fontenc}    
\usepackage{hyperref}       
\usepackage{url}            
\usepackage{booktabs}       
\usepackage{amsfonts}       
\usepackage{nicefrac}       
\usepackage{microtype}      
\usepackage{xcolor}         
\usepackage{amsmath}

\usepackage{graphicx}
\usepackage{wrapfig}
\usepackage{subcaption}
\usepackage{algorithm}
\usepackage{algorithmic}
\usepackage{multirow}
\usepackage{makecell}
\usepackage{natbib}
\setcitestyle{numbers,square}

\title{Beyond Similarity: Personalized Federated Recommendation with Composite Aggregation}

\author{
  Honglei Zhang$^{1,2}$\quad Haoxuan Li$^{3}$ \quad Jundong Chen$^{1,2}$ \quad Sen Cui$^{4}$ \quad Kunda Yan$^{4}$ \\ 
  \textbf{Abudukelimu Wuerkaixi$^{4}$ \quad Xin Zhou$^{5}$ \quad Zhiqi Shen$^{5}$ \quad Yidong Li $^{1,2}$}\thanks{Corresponding author.}\\
  $^1$Key Laboratory of Big Data \& Artificial Intelligence in Transportation, Ministry of Education\\
  $^2$School of Computer Science and Technology, Beijing Jiaotong University\\
  $^3$Peking University \quad $^4$ Tsinghua University \quad $^5$ Nanyang Technological University\\
  \texttt{honglei.zhang@bjtu.edu.cn \quad ydli@bjtu.edu.cn}
}

\begin{document}

\maketitle

\begin{abstract}

Federated recommendation aims to collect global knowledge by aggregating local models from massive devices, to provide recommendations while ensuring privacy. Current methods mainly leverage aggregation functions invented by federated vision community to aggregate parameters from similar clients, e.g., clustering aggregation. Despite considerable performance, we argue that it is suboptimal to apply them to federated recommendation directly. This is mainly reflected in the disparate model architectures. Different from structured parameters like convolutional neural networks in federated vision, federated recommender models usually distinguish itself by employing one-to-one item embedding table. Such a discrepancy induces the challenging \textit{embedding skew} issue, which continually updates the trained embeddings but ignores the non-trained ones during aggregation, thus failing to predict future items accurately. To this end, we propose a personalized \underline{Fed}erated recommendation model with \underline{C}omposite \underline{A}ggregation (FedCA), which not only aggregates \textit{similar} clients to enhance trained embeddings, but also aggregates \textit{complementary} clients to update non-trained embeddings. Besides, we formulate the overall learning process into a unified optimization algorithm to jointly learn the similarity and complementarity. Extensive experiments on several real-world datasets substantiate the effectiveness of our proposed model. The source codes are available at \href{https://github.com/hongleizhang/FedCA}{https://github.com/hongleizhang/FedCA}.

\end{abstract}

\section{Introduction}

Federated recommendation (FR), as an emerging on-device learning paradigm~\cite{yin2024device,singhal2021federated,t2020personalized}, has attracted significant interest from both academia~\cite{fcf_2019,lightfr_2022} and industry~\cite{ning2021learning,hard2018federated}. Existing FRs typically employ different collaborative filtering backbones as their local models~\cite{mf_2009,ncf_2017}, and perform various aggregation functions to obtain a global recommender, following basic federated learning (FL) principles~\cite{fl_first_2017,fedfast_2020}. For instance, one pioneering work is FCF~\cite{fcf_2019}, which is an adaptation of centralized matrix factorization by performing local updates and global aggregation with federated optimization. Besides, FedNCF~\cite{fedncf_2022_kbs} integrates the linearity of matrix factorization with the non-linearity of deep embedding techniques, building upon the foundations of FCF. These embedding-based FR models effectively balance recommendation accuracy and privacy preservation.

Generally, the success of FRs stems from their capability to embody data locality while achieving knowledge globality across multiple clients through aggregation functions~\cite{fcf_2019,fedncf_2022_kbs,li2023federated}. These functions play a crucial role in federated optimization procedures, determining which knowledge from each client and to what extent it is integrated into the global model~\cite{yuan2023hetefedrec}. Among them, the most well-known method is FedAvg, which allocates larger weights to clients with more data samples to achieve weighted aggregation, thus optimizing the global model~\cite{fl_first_2017}. Subsequent works aim to improve aggregation strategies to address the data heterogeneity challenge in federated settings~\cite{liu2021pfa,ji2019learning,ghosh2020efficient}. For instance, PerFedRec first exploits clustering to identify clients with similar data distributions and then conducts group-wise aggregation to accomplish the adaptation~\cite{perfedrec2022}. Besides, FedAtt allocates the attention coefficient of different clients by calculating the similarity between the local and global models, thereby achieving personalized federated optimization~\cite{ji2019learning}. The above aggregation methods effectively mitigate the heterogeneity challenge by considering fine-grained similarity.

Notably, such aggregation functions utilized in FRs are primarily inspired by those in federated vision community, such as weighted aggregation~\cite{fl_first_2017}, clustering aggregation~\cite{liu2021pfa}, and attentive aggregation~\cite{ji2019learning}. All of these are essentially with the similarity assumption, where similar clients are assigned more weights, while dissimilar clients are given relatively smaller weights. Despite achieving satisfactory performance, we argue that directly adopting off-the-shelf aggregation functions from federated vision domain may not be well-adapted to FR tasks, which naturally exhibit significant heterogeneity and are highly required personalization preference for each client.

\begin{wrapfigure}{r}{0.4\textwidth} 
\vspace{-0.4cm}
  \centering
  \includegraphics[width=0.38\textwidth]{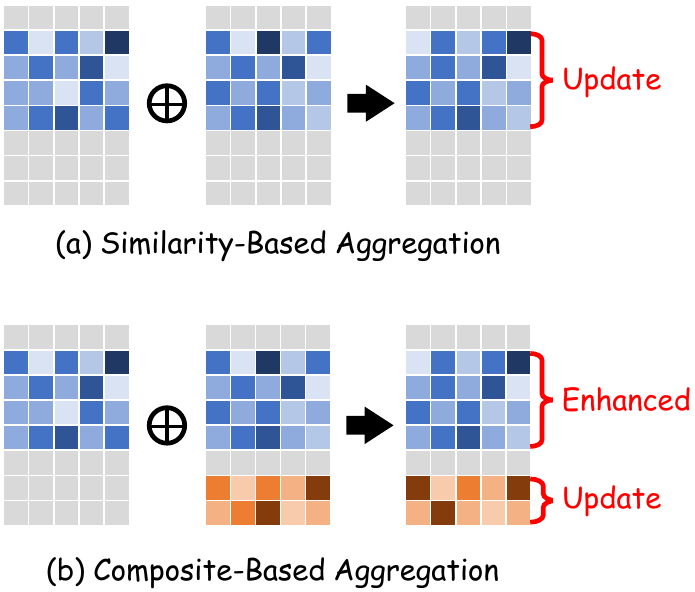}
  \caption{Taking two clients as an ideal example, $\oplus$ denotes aggregation operators. Previous work can only update trained items repeatedly via similarity aggregation, while our composite aggregation method can both enhance trained items and update non-trained items.}
  \label{fig:intro}
  \vspace{-0.6cm}
\end{wrapfigure}

The reason for this research gap is mainly reflected in the disparate model architectures~\cite{xia2022device}. Unlike federated vision models, e.g., convolutional neural networks, typically with a deep network structure (a.k.a., structured parameters), federated recommender models usually distinguish itself by employing one-to-one item embedding table. Since different clients may involve distinct subsets of interacted items, leading to different rows trained in the embedding table for each client. When only relying on similarity aggregation, it leads to the unique \textit{embedding skew} issue in FRs, where trained embeddings (blue) continually improve while non-trained embeddings (grey) keep intact or even deteriorate during aggregation, as depicted in Fig.\ref{fig:intro} (a). Hence, it poses a great challenge to predict uninteracted items in local device solely by similarity aggregation.

In this work, we take the first step in exploring aggregation mechanisms for FR models and identify the unique embedding skew issue in FR tasks. In light of this, we propose a composite aggregation mechanism tailored to FR models, which aggregates not only similar clients but also complementary ones. Such a mechanism can enhance the already trained embeddings, and update those that were not trained, thus enhancing the ability to predict future items on edge devices, as shown in Fig.\ref{fig:intro} (b). Besides, we formulate the aggregation process into a unified optimization algorithm to jointly learn the similarity and complementarity. Extensive experiments on several real-world datasets show that our model consistently outperforms several state-of-the-art methods.


\section{Related Work}

\vspace{-0.2cm}

\textbf{Traditional FRs} aim to learn a shared item representation for all clients and a private user representation for each client. It mainly comprises three modules: a private user encoder, a shared item encoder, and a fusion module~\cite{yin2024device,zhang2023dual,hpfl_2021}, following basic FL principle~\cite{fl_first_2017,fedprox_2020,fedrep_2021}. Some attempts are launched to follow these three research lines~\cite{hpfl_2021,lightfr_2022,fedncf_2022_kbs}. Specifically, HPFL introduced a hierarchical user encoder to differentiate between private and public user information~\cite{hpfl_2021}. Zhang \textit{et al.} proposed federated discrete optimization model to learn binary codes of item encoder~\cite{lightfr_2022}. FedNCF attempted to use a multi-layer perceptron fusion module to learn non-linearities between users and items~\cite{fedncf_2022_kbs}.

To achieve \textbf{personalized FRs}, some pioneer works aim to learn personalized item encoders~\cite{zhang2023dual,hanzely2020lower,li2023federated}, such as dual personalization~\cite{zhang2023dual} and additive personalization~\cite{li2023federated}. Note that the above methods employ the classic FedAvg for aggregation~\cite{fl_first_2017}. Subsequent methods attempt to improve the effectiveness of FedAvg, such as clustering~\cite{fedfast_2020,ghosh2020efficient,perfedrec2022, co_cluster_2024}, attention~\cite{ji2019learning} and graph aggregation~\cite{ye2023personalized}. For instance, FedFast~\cite{fedfast_2020} used clustering aggregation to enhance training efficiency, while FedAtt~\cite{ji2019learning} utilized attention to learn coefficients between global and local models. pFedGraph~\cite{ye2023personalized} introduced a collaborative graph to learn the similarity between individuals. Despite achieving considerable results, all aggregation methods are with similarity assumption, which is more suitable for structured parameters in federated vision tasks. Different from previous work, we are the first work to design composite aggregation mechanism tailored for FR tasks, which simultaneously considers similarity and complementarity to more effectively aggregate embedding tables.

\section{Empirical Analysis}\label{sec:emp}


\begin{wrapfigure}{r}{0.4\textwidth}
\vspace{-0.4cm}
\begin{minipage}{\linewidth}
  \centering
  \begin{subfigure}{0.45\linewidth}
  \centering
    \includegraphics[width=\linewidth]{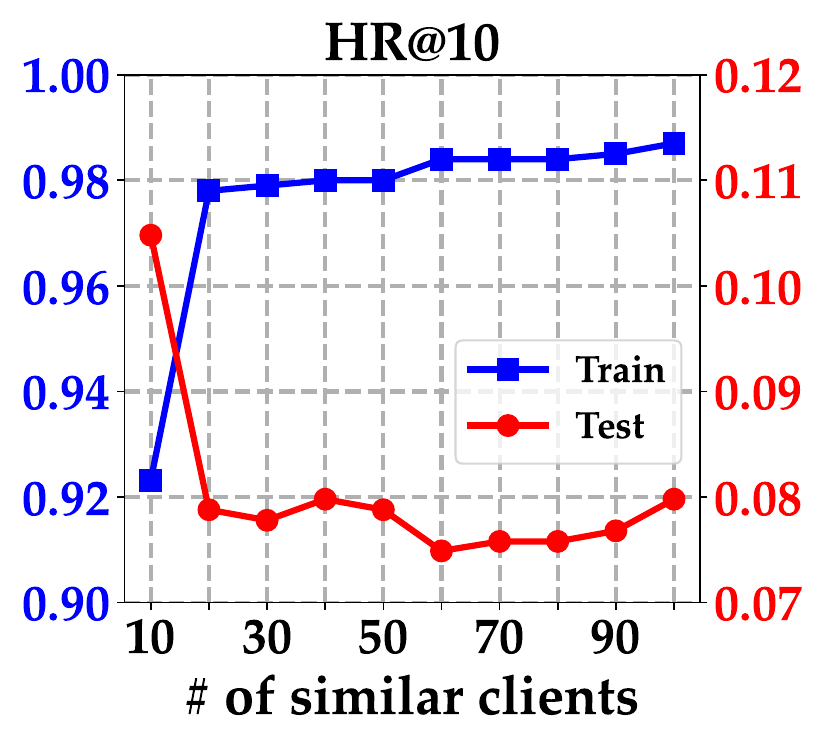}
   \end{subfigure}
     \begin{subfigure}{0.45\linewidth}
  \centering
    \includegraphics[width=\linewidth]{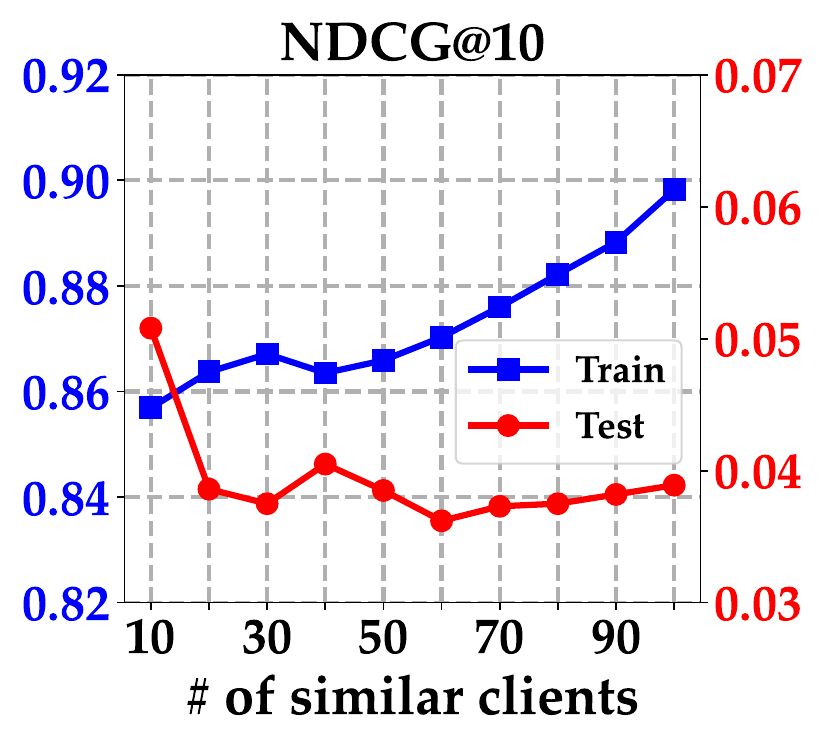}
   \end{subfigure}
\vspace{-0.2cm}
    \caption*{(a) Filmtrust}
        \begin{subfigure}{0.45\linewidth}
  \centering
    \includegraphics[width=\linewidth]{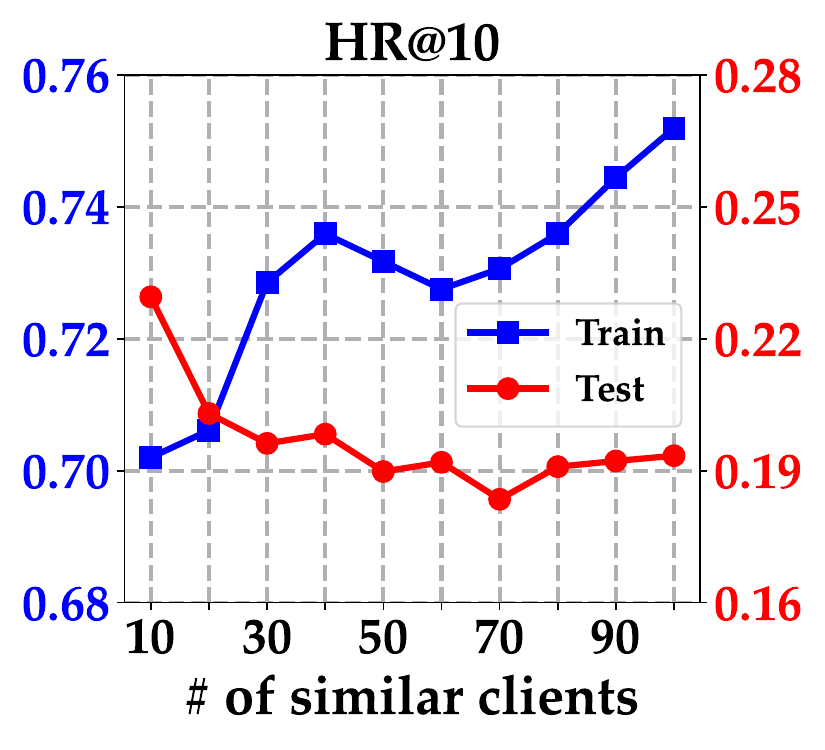}
   \end{subfigure}
        \begin{subfigure}{0.45\linewidth}
  \centering
    \includegraphics[width=\linewidth]{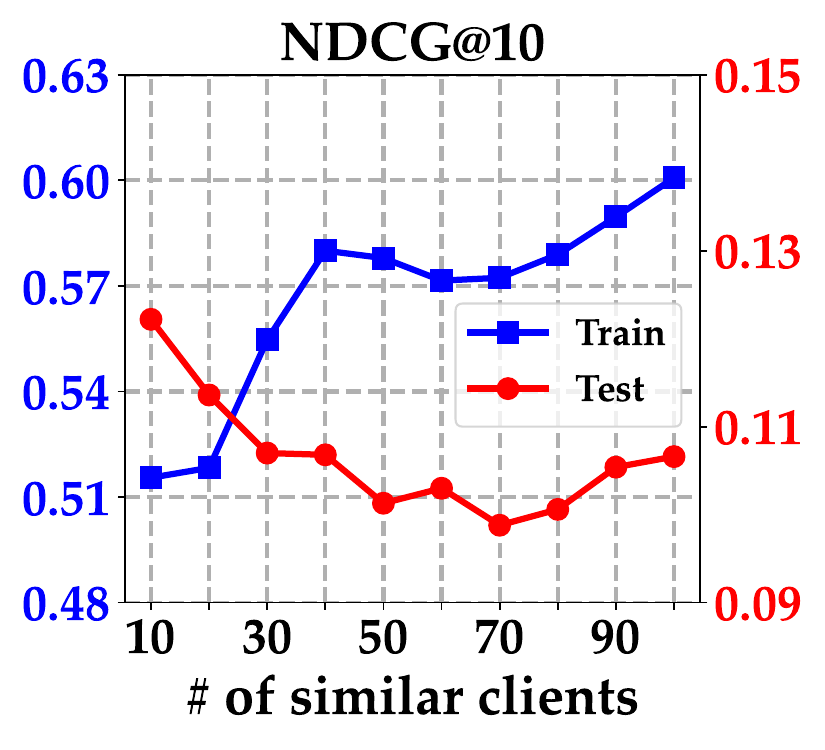}
   \end{subfigure}
   \vspace{-0.2cm}
    \caption*{(b) Movielens}
    \caption{Empirical results regarding HR@10 and NDCG@10 on train and test sets, respectively.}
    \label{fig:emp}
\end{minipage}
\vspace{-0.6cm}
\end{wrapfigure}

By analyzing disparate model architectures with federated vision models, we intuitively explored the embedding skew issue that uniquely occurred during aggregation process in FR tasks. To experimentally validate our findings, as illustrated in Fig.\ref{fig:intro} (a), this section conducts verification analysis on two commonly used datasets (Filmtrust~\cite{filmtrust_2013} and Movielens~\cite{movielens_2015}) in FR tasks, aiming to show the unique embedding skew issue from an empirical perspective.

Specifically, we conduct exploratory experiments with improved FedAvg model by aggregating parameters with $s\in\{10, 20, \cdots, 100\}$ most similar clients. As depicted in Fig.\ref{fig:emp}, it is evident that as the number of aggregated similar clients increases, the accuracy (HR@10 and NDCG@10) of the train set continues to rise, while that of the test set generally declines on both datasets. Typically, the accuracy trends of the training set and the test set should be consistent. This ultimately results in a widening gap between the train and test sets, indicative of performance degradation caused by embedding skew, i.e., trained embeddings of interacted items greatly enhanced while untrained embeddings of non-interacted items deteriorated during aggregation. The observation indicates that solely utilizing similarity to aggregate embedding tables is suboptimal. This also aligns with the motivation behind our proposed model, which exploits composite aggregation considering both similarity and complementarity. Hence, it can delicately improve generalization on test sets, thus enabling accurate recommendations of future items on edge devices.



\section{Problem Formulation}


Here we introduce the basic notations, general FR framework and rethinking heterogeneity in FRs.

\textbf{Notations.} Assume there are $n$ users/clients $\mathcal{U} = \{u\}$, and $m$ items $\mathcal{I} = \{i\}$ stored in the server. Each user $u$ keeps a local dataset $\mathcal{D}_u$, which comprises tuples $(u,i,r_{ui} | i\in\mathcal{I}_u)$, where $\mathcal{I}_u$ denotes the observed items for client $u$, and each entry $r_{ui}\in \{0,1\}$ indicates the label for user $u$ on item $i$. The goal of FRs is to predict $\hat{r}_{ui}$ of user $u$ for each future item $i \in \mathcal{I} \setminus \mathcal{I}_u$ on local devices.


\textbf{General FR Framework.} Formally, the global objective of general FR framework over $n$ clients is
\begin{equation}\label{eq:frs_fl}
\min _{\left(\mathbf{p}_1, \mathbf{p}_2,\cdots, \mathbf{p}_n; \Theta_1, \Theta_2,\cdots, \Theta_n\right)} \frac{1}{n} \sum_{u=1}^n p_u \mathcal{L}_u\left(\mathbf{p}_u, \Theta_u ; \mathcal{D}_u\right),
\end{equation}
\noindent where $\mathbf{p}_u$ and $\Theta_u$ denote local user embedding and global parameters, respectively. The server aggregates $\Theta_u$ with aggregation weight $p_u$,  e.g., $p_u=|\mathcal{D}_u|/|\mathcal{D}|$ in FedAvg~\cite{fl_first_2017} to facilitie global update. $\mathcal{L}_u$ is the task-specific objective (e.g., log loss~\cite{ncf_2017}) to facilitie local training. Traditional FRs attempt to learn a global model $\Theta$ across $n$ clients, where $\Theta=\Theta_1=\cdots=\Theta_n$~\cite{fcf_2019,fedncf_2022_kbs}, while personalized FRs keep different local models $\Theta_u$ to achieve high efficacy on their local clients~\cite{li2023federated,zhang2023dual}. Note that unless otherwise specified, we will use item embedding table $\mathbf{Q}_u$ to instantiate general $\Theta_u$ in following sections, since $\mathbf{Q}_u$ is the standard configuration in embedding-based FRs~\cite{fcf_2019,fedfast_2020}.

\textbf{Rethinking heterogeneity in FRs.} The FR tasks naturally exhibit great heterogeneity in each client, since the item sets interacted by each are vastly different, radically causing the embedding skew issue. Formally, we assume $p(x,y)$ to denote the joint distribution of features $x\in\{(u,i) | i\in\mathcal{I}_u\}$ and labels $y\in\{r_{ui}|i\in\mathcal{I}_u\}$. For two heterogeneous clients $u$ and $v$, it is evident that $p(x^u,y^u)\neq p(x^v,y^v)$. By decomposing the joint distribution $p(x, y)=p(x)p(y|x)$, heterogeneity can be represented as
\begin{equation}\label{eq:heterogeneity}
p\left(x^{u}\right) p\left(y^{u} | x^{u}\right) \neq p\left(x^{v}\right) p\left(y^{v} | x^{v}\right).
\end{equation}
Currently, some methods utilize model similarity to mitigate the \textit{concept shift} problem~\cite{feddrift}, by aggregating similar conditional distributions to ensure that $p\left(y^{u} | x^{u}\right) \approx p\left(y^{v} | x^{v}\right)$~\cite{liu2021pfa,ghosh2020efficient}. Building on this, if we aim to ensure $p(x^u,y^u)\neq p(x^v,y^v)$, then $p(x^u)\neq p(x^v)$ should be explicitly satisfied. This precisely reflects the complementarity of local data. Thus we propose a composite aggregation mechanism to ensure both model similarity and data complementarity, aiming to more accurately model the fine-grained heterogeneity between the joint distribution $p(x^u,y^u)$ and $p(x^v,y^v)$.





\section{The Proposed FedCA Model}

In this section, we elaborate on our proposed framework personalized \underline{Fed}erated recommendation with \underline{C}omposite \underline{A}ggregation (FedCA), which considers both model similarity and data complementarity. The goal is to alleviate the embedding skew issue inherit in FR tasks. Concretely, we first formulate a unified learning framework to optimize similarity and complementarity. Then, we provide a detailed optimization for server aggregation, followed by the local training and inference process on the client-side. Finally, we discuss the relationship between FedCA and other aggregation mechanisms.

\textbf{The Overall Learning Framework.} From a global perspective, we integrate the server aggregation and local training into a unified optimization framework tailored for FR tasks, as shown in Eq.(\ref{eq:fedca}). It aims to optimize the personalized local parameters $\{\mathbf{p}_u,\mathbf{Q}_u\}$ and aggregation weight vector $\{\mathbf{w}_u\}$ for each client, which is influenced by the joint constraints of similarity and complementarity.
\begin{align}\label{eq:fedca}
\min_{\{\mathbf{p}_u,\mathbf{Q}_u,\mathbf{w}_u\}} \sum_{u=1}^n & \left(\mathcal{L}_u(\mathbf{p}_u, \mathbf{Q}_u; \mathcal{D}_u) + \alpha\sum_{v=1}^n \mathcal{F}_s(w_{uv};\mathbf{Q}_u;\mathbf{Q}_v)  + \beta\sum_{v=1}^n \mathcal{F}_c(w_{uv};\mathcal{D}_u;\mathcal{D}_v) \right) \\
& \text{s.t.} \quad \mathbf{1}^T \mathbf{w}_u = 1; \,\,\, \mathbf{w}_u > \mathbf{0}. \nonumber
\end{align}
\noindent where the term $\mathcal{L}_u(\cdot)$ denotes the local empirical risk towards model parameters $\mathbf{p}_u$ and $\mathbf{Q}_u$, following the weighted aggregation of model parameters $\mathbf{Q}_u=\sum_{v=1}^n w_{nv}\mathbf{Q}_v$ downloaded from the server.  The term $\mathcal{F}_s(\cdot)$ represents the model similarity between $\mathbf{Q}_u$ and $\mathbf{Q}_v$, while the term $\mathcal{F}_c$ quantifies the data complementarity between $\mathcal{D}_u$ and $\mathcal{D}_v$. The two constraints ensure $\mathbf{w}_u$ satisfy normalization and non-negativity. Besides, $\alpha$ and $\beta$ are tuning coefficients. Through the unified learning framework, we jointly optimize $\mathbf{w}_u$ to a balance point to suitably aggregate item embeddings, thereby considering both similarity and complementarity during the server aggregation and local training procedures.

\textbf{Server Aggregation.} The server's responsibility is to optimize the aggregation weight $\mathbf{w}_u$ for each client $u$, thus achieving personalized global aggregation for each client. Ideally, we aim for $\mathbf{w}_u$ to be perfectly optimized under the loss function in Eq.(\ref{eq:fedca}). However, this is impractical due to constraints imposed by the federated setting. The server can only access the local models $\mathbf{Q}_u$ uploaded by each client, without detailed knowledge of each client's user embedding $\mathbf{p}_u$ and local data $\mathcal{D}_u$, thus making it challenging to directly compute $\mathcal{L}_u$ at the server side. To reasonably perceive initial contribution of each client, we utilize the mean squared loss between $\mathbf{w}_{u}$ and relative quantity of local data $\mathbf{p}$ as a proxy for $\mathcal{L}_u$, measuring the optimization level of each client, inspired by recent work~\cite{ye2023personalized}. Therefore, the loss function to optimize $\mathbf{w}_u$ at the server side is rewritten as Eq.(\ref{eq:optimize_w}).
\begin{align}\label{eq:optimize_w}
\min_{\mathbf{w}_u} \sum_{v=1}^n & \left((w_{uv}-p_v)^2 + \alpha \mathcal{F}_s(w_{uv};\mathbf{Q}_u;\mathbf{Q}_v)  + \beta \mathcal{F}_c(w_{uv};\mathcal{D}_u;\mathcal{D}_v) \right) \\
& \text{s.t.} \quad \mathbf{1}^T \mathbf{w}_u = 1; \,\,\, \mathbf{w}_u > \mathbf{0}. \nonumber
\end{align}
The above formulation minimizes the pre-defined supervised loss while balancing the similarity and complementarity. To measure the similarity between conditional distributions of clients $p(y|x)$, we adopt common practices~\cite{feddrift}, i.e., local model parameters to capture the mapping from the marginal distribution $p(x)$ to the label distribution $p(y)$. Hence the term $\mathcal{F}_s$ can be represented as in Eq.(\ref{eq:func_s}).
\begin{align}\label{eq:func_s}
\mathcal{F}_s(w_{uv};\mathbf{Q}_u;\mathbf{Q}_v)= \left(w_{uv}-\sigma(\mathbf{Q}_u,\mathbf{Q}_v) \right)^2
\end{align}
\noindent where $\sigma(\cdot)$ denotes the similarity function and here $\sigma(\mathbf{Q}_u, \mathbf{Q}_v)=1/(1+\parallel\mathbf{Q}_u-\mathbf{Q}_v\parallel^2)$. It can be switched to any similarity function, e.g., cosine similarity. The function $\mathcal{F}_s$ ensures that the aggregation weight $w_{uv}$ increases when the models of two clients are highly similar. To assess the complementarity of client data about marginal distributions $p(x)$ at the server side, we utilize the intermediate features as proxies for local data $\mathcal{D}_u$, i.e., the subset of item embeddings $\mathbf{Q}_u^s$ corresponding to the local interacted item sets $\mathcal{I}_u$. To further guard against input reconstruction attacks in FRs~\cite{lightfr_2022}, we perform Singular Value Decomposition (SVD) on $\mathbf{Q}_u^s$ and then retain the left singular matrix with first $k$ columns. This yields a privacy-enhanced representation $\mathbf{X}_u$ of the local data. Inspired by mutual information theory~\cite{mutual_2004}, the term $\mathcal{F}_c$ can be represented as in Eq.(\ref{eq:func_c}).
\begin{align}\label{eq:func_c}
\mathcal{F}_c(w_{uv};\mathcal{D}_u;\mathcal{D}_v)= -w_{uv}\cdot\cos (  \phi(\mathbf{X}_u,\mathbf{X}_v) ) ,
\end{align}
\noindent where $\phi(\mathbf{X}_u,\mathbf{X}_v)=\frac{1}{k}\sum_{l=1}^k \arccos({\mathbf{x}_u^l}^T \mathbf{x}_v^l)$. $\phi(\cdot)$ is used to measure the angle between the $l$-th singular vectors corresponding to the data of two clients. When the lengths of two vectors are unequal, we apply a padding operation to keep consistency in lengths. The function $\mathcal{F}_c$ ensures that when the angle between two clients is orthogonal, the smaller the mutual information between $\mathbf{X}_u$ and $\mathbf{X}_v$, implying a great complementarity between the two clients. We denote the similarity vector computed by $\sigma(\cdot)$ for each user $u$ as $\mathbf{s}_u$ and the complementarity vector computed by $\phi(\cdot)$ as $\mathbf{c}_u$. We can see that Eq.(\ref{eq:optimize_w}) can be easily rewritten as a standard quadratic program problem regarding the aggregation weight $\mathbf{w}_u$. Hence, it can be efficiently solved by classic convex optimization solvers~\cite{diamond2016cvxpy}.

By introducing both similarity and complementarity into the overall learning framework, we can not only aggregate similar item embeddings but also aggregate complementary ones, thereby alleviating the embedding skew issue. Importantly, it can ensure consistency in conditional distributions $p(y|x)$ while preserving complementarity in marginal distributions $p(x)$ to better model client heterogeneity by the joint distributions $p(x,y)$. Besides, we can conclude that our proposed composite aggregation mechanism only requires efficient computation on the powerful server side, without extra computational overhead to the resource-constrained clients. Its computational complexity on local devices is consistent with classical FR methods~\cite{fcf_2019,fedncf_2022_kbs}, thereby enabling efficient on-device training.

\textbf{Local Training.} The mission of each client $u$ is to utilize local data to optimize the local empirical loss $\mathcal{L}_u$ regarding private user embedding $\mathbf{p}_u$ and personalized item embedding $\mathbf{Q}_u$. The private user embedding $\mathbf{p}_u$ are kept locally, while the computed item embedding $\mathbf{Q}_u$ is uploaded to the server for global aggregation. To mine the information from interactions during training, we specify $\mathcal{L}_u$ as the binary cross-entropy (BCE) loss, which is a well-designed objective function for recommender systems. Formally, the objective function of BCE loss is defined in Eq.(\ref{eq:bce}).
\begin{align}\label{eq:bce}
\mathcal{L}_{u}=-\sum_{(u, i) \in \mathcal{D}_u} r_{u i} \log \hat{r}_{u i}+\left(1-r_{u i}\right) \log \left(1-\hat{r}_{u i}\right),
\end{align}
\noindent where $\mathcal{D}_u=\mathcal{D}_u^{+} \cup \mathcal{D}_u^{-}$ and $\mathcal{D}_u^{+}$ represents observed interactions, i.e., $r_{ui} = 1$, and $\mathcal{D}_u^{-}$ represents uniformly sampled negative instances, i.e., $r_{ui} = 0$. Note that unlike federated vision tasks, which require the proximal term to restrict personalized models to be closer to the global model, i.e., $\parallel \mathbf{Q}_u-\mathbf{Q}_g \parallel^2$ in FedProx~\cite{fedprox_2020} and pFedGraph~\cite{ye2023personalized}, etc., where $\mathbf{Q}_g$ denotes the global model, FR tasks inherently involve great heterogeneity and strong requirements for personalization. Hence, we solely use task-driven loss $\mathcal{L}_u$ without additional terms to keep the localization properties of item embedding. By optimizing the BCE loss in the local client, we can update the user embedding $\mathbf{p}_u$ and $\mathbf{Q}_u$ by stochastic gradient descent as follows:
\begin{align}\label{eq:local_train}
\mathbf{p}_u=\mathbf{p}_u-\eta \cdot \frac{\partial \mathcal{L}_{u}}{\partial \mathbf{p}_u},\quad \mathbf{Q}_u=\mathbf{Q}_u-\eta \cdot \frac{\partial \mathcal{L}_{u}}{\partial \mathbf{Q}_u},
\end{align}
\noindent where $\eta$ is the learning rate. At the end of local training in each round, clients upload their local item embeddings $\mathbf{Q}_u$ to the server for global aggregation. 

\textbf{Local Inference.} During the local inference stage, client $u$ first downloads the aggregated item embeddings $\mathbf{Q}_g = \sum_{v=1}^n w_{uv}\mathbf{Q}_v$ from the server. Notably, in federated vision domains, it can directly perform local inference using global parameters $\mathbf{Q}_g$. However, in FR tasks, the existence of client-specific user embedding $\mathbf{p}_u$ introduces a spatial misalignment issue between the user embedding $\mathbf{p}_u^{t-1}$ at previous round $t-1$ and the aggregated item embedding $\mathbf{Q}_g^t$ at this round $t$. To achieve space alignment, we employ a simple yet effective interpolation method to narrow the gap between local-specific parameters $\mathbf{p}_u$ and global parameters $\mathbf{Q}_g$, i.e., $\mathbf{Q}_u^t=\rho\mathbf{Q}_u^{t-1}+(1-\rho)\mathbf{Q}_g^{t}$ where $\rho$ controls the weight of the local parameters in the current round. By introducing $\rho$, we balance the weight of local parameters $\mathbf{Q}_u$ and global aggregated parameters $\mathbf{Q}_g$, thereby aligning items with users in the embedding space. After obtaining the item embedding $\mathbf{q}_u^i\in\mathbf{Q}_u$ for each item $i$, we can perform local inference with $\mathbf{p}_u$ at the local client $u$, which is $\hat{r}_{u i}=f\left(\mathbf{p}_u, \mathbf{q}_u^i\right)$, where $f(\cdot)$ denotes the inner product or neural match function~\cite{ncf_2017} to compute similarities between user $u$ and item $i$. By aggregating both similar and complementary clients, our model can enhance the prediction accuracy for future items. We present the FedCA algorithm in detail in Algorithm~\ref{alg:fedca}.

\begin{wrapfigure}{r}{0.55\textwidth}
\vspace{-0.7cm}
\begin{minipage}{1.0\linewidth}
\begin{algorithm}[H]
\textbf{Input:} local models: \(\mathbf{p}_u,\mathbf{Q}_u, \mathbf{w}_u;\) global rounds: \(T\) \\
local epochs: \(E\); learning rate \(\eta\); selected clients $\mathcal{U}_s$;\\
\textbf{Server executes:} \\
\vspace{-0.4cm}
\caption{FedCA}
\label{alg:fedca}
\begin{algorithmic}[1]
\STATE Initialize global item embeddings $\{\mathbf{Q}_u\}_{u=1}^n$;
\FOR{each round $t=1,2,\cdots, T$}
    \STATE Sends $\mathbf{Q}_u^t=\sum_{v=1}^n w_{uv}\mathbf{Q}_v^t$ to each client $u$;
    \FOR{each client $u\in \mathcal{U}_s$ \textbf{in parallel}}
        \STATE $\mathbf{Q}_u^{t+1}\leftarrow$ LocalTraining($\mathbf{Q}_u^t$, $u$);
    \ENDFOR
    \STATE $\mathbf{s}_u \leftarrow$ compute similarity with Eq.(\ref{eq:func_s});
    \STATE $\mathbf{c}_u \leftarrow$ compute complementarity with Eq.(\ref{eq:func_c});
    \STATE $\mathbf{w}_u \leftarrow$ optimize with Eq.(\ref{eq:optimize_w}) for each client $u$;    
\ENDFOR
\end{algorithmic}

{\bf LocalTraining}$(\mathbf{Q}_u, $u$)$: \\
\vspace{-0.4cm}
\begin{algorithmic}[1]
\FOR{each local epoch $e=1,2,\cdots, E$}
    \FOR{each batch in $\mathcal{D}_u$}
        \STATE compute local loss $\mathcal{L}_u$ by following Eq.(\ref{eq:bce});
        \STATE update $\mathbf{p}_u$ and $\mathbf{Q}_u$ with Eq.(\ref{eq:local_train});
    \ENDFOR
\ENDFOR
\RETURN $\mathbf{Q}_u$
\end{algorithmic}
\end{algorithm}
\end{minipage}
\vspace{-0.2cm}
\end{wrapfigure}

\textbf{Relations with Classic Aggregation Mechanisms.} In this section, we will analyze the compatibility of our proposed FedCA model. We introduce a unified optimization framework for aggregating item embeddings in FR tasks. This framework can transform into several classical aggregation methods by flexibly adjusting hyperparameters $\alpha$ and $\beta$, as well as the proxy coefficient $p_u$. Specifically, when $\alpha=0$ and $\beta=0$, and the proxy coefficient $p_u$ is set to the mean, our method degrades to the average aggregation method used in FCF~\cite{fcf_2019}. When $\alpha=0$ and $\beta=0$, and $p_u$ is set to the relative dataset size, our method achieves the weighted aggregation used in FedAvg~\cite{fl_first_2017}. When $\alpha=0$ and $\beta=0$, and $p_u$ is set to the degree of difference between local and global models, our method can degrade to the FedAtt method~\cite{ji2019learning}. When $\alpha \neq 0$ and $\beta=0$, it can become the similarity-based aggregation method pFedGraph~\cite{ye2023personalized}. Specifically, if only the most similar clients are selected for each client, it is equivalent to the clustering aggregation in PerFedRec~\cite{perfedrec2022}. When $\alpha=0$ and $\beta \neq 0$, it implies aggregating only dissimilar parameters for each client, which is equivalent to the FedFast~\cite{fedfast_2020}, where clients are first clustered, and then clients from each cluster are aggregated proportionally. We conclude that our method can flexibly implement several aggregation methods.

\textbf{Privacy Discussion.} Our FedCA method maintains the same privacy protection standards as the baselines since it requires transmitting model parameters or gradients and does not share any raw data with third parties. Besides, since our composite aggregation mechanism is model-agnostic, it can seamlessly integrate with other privacy-enhanced FR models like FMF-LDP~\cite{minto2021stronger} and FR-FMSS~\cite{li2023federated}, and can easily incorporate various advanced privacy protection strategies, such as differential privacy~\cite{dp_2021} and label flipping~\cite{labelflip_2020}, to further enhance user privacy guarantees.

\section{Experiments}\label{sec:main_exp}

In this section, we provide detailed experimental settings and comprehensive experimental results.

\subsection{Experimental Settings}

\textbf{Datasets.} We evaluate our model on four benchmark datasets with varying client scales: Movielens-100K (ML-100K)~\cite{movielens_2015}, Filmtrust~\cite{filmtrust_2013}, Movielens-1M (ML-1M)~\cite{movielens_2015}, and Microlens-100K (MC-100K)~\cite{microlens_2023}. The first three datasets are for movie recommendation with explicit feedback, where ratings greater than 0 are converted to 1. The last dataset is for short video recommendation with implicit feedback. Each user is treated as an independent client, and each client's data inherently exhibits great heterogeneity.

\textbf{Baselines.} To thoroughly explore the effectiveness of various aggregation mechanisms, we compared 8 classic federated models, including: 1) \textit{Local:} local training without federated aggregation. 2) \textit{FCF:}~\cite{fcf_2019} averaged aggregation by allocating equal weights to each client. 3) \textit{FedAvg:}~\cite{fl_first_2017} weighted aggregation by the relative size of local client data. 4) \textit{PerFedRec:}~\cite{perfedrec2022} clustering aggregation by grouping clients into several clusters with model similarities. 5) \textit{FedAtt:}~\cite{ji2019learning} attentive aggregation by minimizing the weighted distance between global and local models.  6) \textit{FedFast:}~\cite{fedfast_2020} active aggregation by identifying representatives from different clusters. 7) \textit{pFedGraph:}~\cite{ye2023personalized} graph aggregation by learning the similarities between individuals. 8) \textit{PFedRec:}~\cite{zhang2023dual} recent personalized FR model, achieved through dual personalization of score function and item embedding.

\textbf{Implementations.} Following previous works~\cite{li2023federated}, we randomly sample $N=4$ negative instances for each positive sample, and utilize the leave-one-out strategy for efficient validation. For our model, we set $k=4$ and choose the optimal values for $\rho\in[0,1]$, $\alpha\in[0,1]$ and $\beta\in[0,1]$ for four different datasets. Besides, we utilize two common evaluation metrics for item ranking tasks: HR@K and NDCG@K where $K=10$. We conduct hyperparameter tuning for all compared models and report the results as the average of 5 repeated experiments. To validate the model agnostics of our method, we verify its effectiveness on two canonical backbones, PMF~\cite{pmf_2007} and NCF~\cite{ncf_2017}.

\begin{table*}[htbp!]
\renewcommand\arraystretch{1.1}
\caption{Comparison results of FedCA and other baselines evaluated on four commonly used datasets. Higher values indicate better performance. The best results are in bold.}
\resizebox{1.0\textwidth}{!}{
\begin{centering}
  \begin{tabular}{c|l|cc|cc|cc|cc}
    \toprule
     \multirow{2}{*}{{Backbones}} & \multirow{2}{*}{{ Models}} &  \multicolumn{2}{c}{ML-100K}  & \multicolumn{2}{c}{Filmtrust}  & \multicolumn{2}{c}{ML-1M} & \multicolumn{2}{c}{MC-100K}   \\ \cline{3-10}
        &  & {HR@10}  & {NDCG@10}  & {HR@10}  & {NDCG@10}  & {HR@10}  & {NDCG@10}  & {HR@10}  & {NDCG@10}   \\
    \midrule
     \multirow{7}*{PMF}  & Local & 0.4128  & 0.2203  & 0.4760 & 0.2410  & 0.4264 & 0.2314 & 0.1246  & 0.0567  \\
     & FCF  &  0.4327  & 0.2497  & 0.6407 & 0.4914  & 0.4454 & 0.2484 & 0.1294  & 0.0594  \\
     & FedAvg  & 0.4878  & 0.2786 & 0.6517 & 0.5126  & 0.4912 & 0.2751 & 0.1295 & 0.0601 \\ 
     & PerFedRec & 0.4973  & 0.2797  & 0.6577  & 0.5247  & 0.4623  & 0.2622  & 0.1255 & 0.0599   \\
     & FedAtt  &  0.4645  & 0.2558  & 0.6088 & 0.3359  & 0.4310 & 0.2168 & 0.0982  & 0.0445  \\
    & FedFast  & 0.4984    & 0.2747  & 0.6527  & 0.5191   & 0.5061 & 0.2898  & 0.1278 & 0.0600 \\
     & pFedGraph  &  0.5928  & 0.4025  & 0.6961 & 0.5430  & 0.7904 & 0.6347 & 0.1324  & 0.0605  \\
    & PFedRec   &  0.7254  & 0.4648  & 0.7096  & 0.5629  & 0.8032 & 0.6519  & 0.1334 & 0.0621   \\
    & FedCA  & \textbf{0.8738}  & \textbf{0.7597}  &  \textbf{0.7725} & \textbf{0.5945}  & \textbf{0.8348}  & \textbf{0.7118}   &\textbf{0.1351}  &\textbf{0.0678}  \\
    \hline
    \multirow{7}*{NCF}  & Local & 0.4077  & 0.2145  & 0.4312 & 0.2485  & 0.3881 & 0.1839 & 0.1004  & 0.0459  \\
    & FCF  &  0.4115  & 0.2390  & 0.6477 & 0.4968  & 0.4269 & 0.2232 & 0.1290  & 0.0668  \\
     & FedAvg  & 0.4478  & 0.2731 & 0.6507 & 0.4969  & 0.4899 & 0.2703 & 0.1397 & 0.0674 \\ 
     & PerFedRec & 0.4135  & 0.2253  & 0.3752  & 0.1418  & 0.4219  & 0.2093  & 0.1128 & 0.0591   \\
     & FedAtt  &  0.4910  & 0.2626  & 0.6547 & 0.4801  & 0.4136 & 0.2177 & 0.1375  & 0.0669  \\
    & FedFast  & 0.4436    & 0.2708  & 0.6632  & 0.5007   & 0.4040 & 0.2008  & 0.1402 & 0.0774 \\
     & pFedGraph  &  0.5822  & 0.3587  & 0.6718 & 0.5021  & 0.5113 & 0.2992 & 0.1416  & 0.0669  \\
    & PFedRec   &  0.6931  & 0.5031  & 0.6732  & 0.5031 & 0.6826 & 0.4041  & 0.1422 & 0.0687    \\
    & FedCA  & \textbf{0.8452}  & \textbf{0.7444}  &  \textbf{0.6836} & \textbf{0.5099}  & \textbf{0.7815}  & \textbf{0.6662}   &\textbf{0.1465}  &\textbf{0.0782}  \\
    \bottomrule
  \end{tabular}
  \end{centering}
  }
  \label{tab:main_results}
  \vspace{-0.2cm}
\end{table*}

\subsection{Experimental Results}

This section introduces the effectiveness of our method through various experiments, including overall performance, analyses with different ratios of training data, and visualization results.


\begin{wrapfigure}{r}{0.5\textwidth}
\vspace{-0.6cm}
\begin{minipage}{\linewidth}
  \centering
  \begin{subfigure}{0.45\linewidth}
  \centering
    \includegraphics[width=\linewidth]{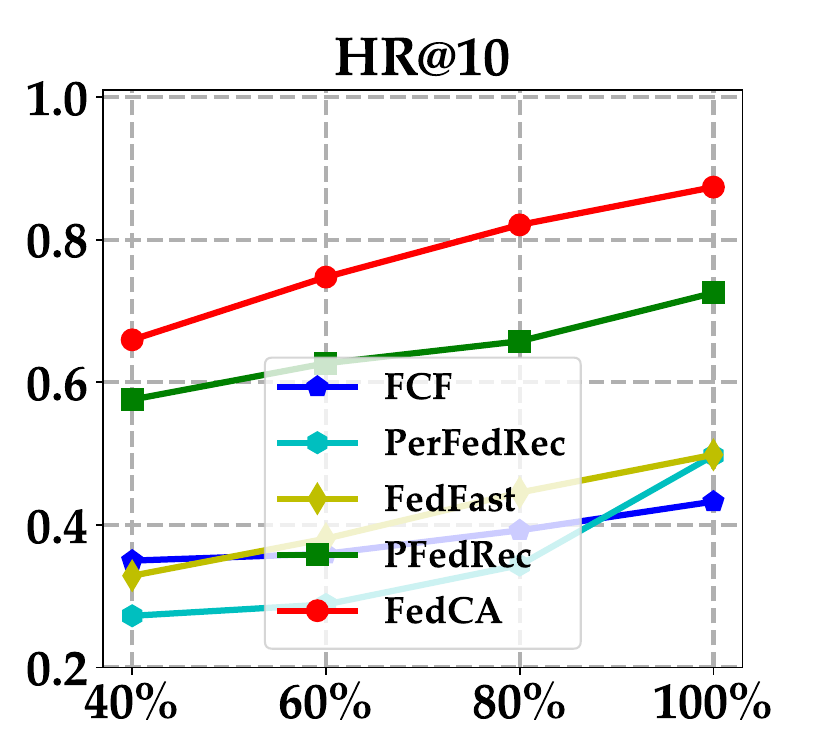}
    \caption*{(a) ML-100K}
   \end{subfigure}
     \begin{subfigure}{0.45\linewidth}
  \centering
    \includegraphics[width=\linewidth]{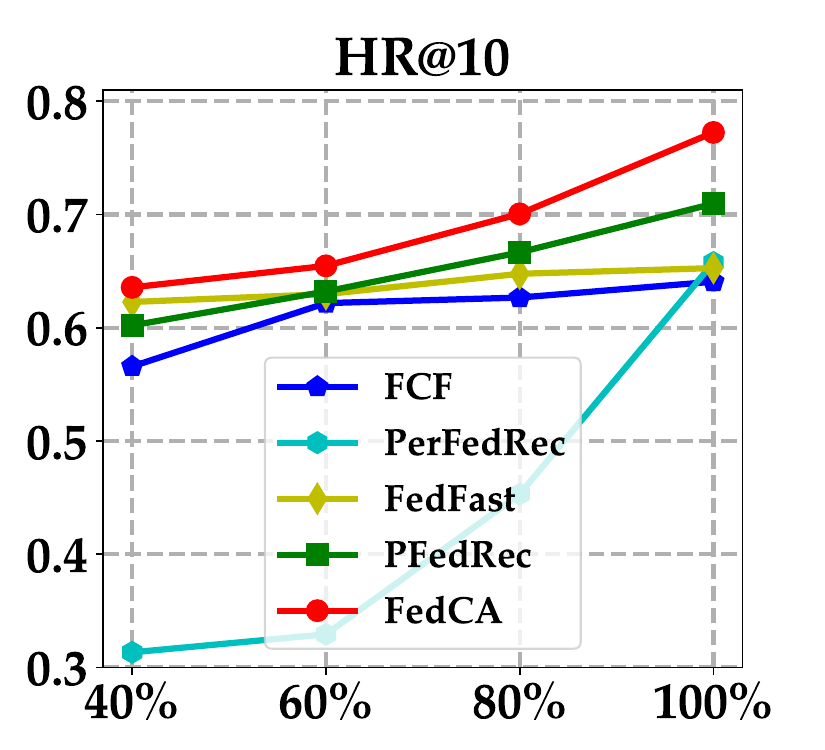}
    \caption*{(b) Filmtrust}
   \end{subfigure}
        \begin{subfigure}{0.45\linewidth}
  \centering
    \includegraphics[width=\linewidth]{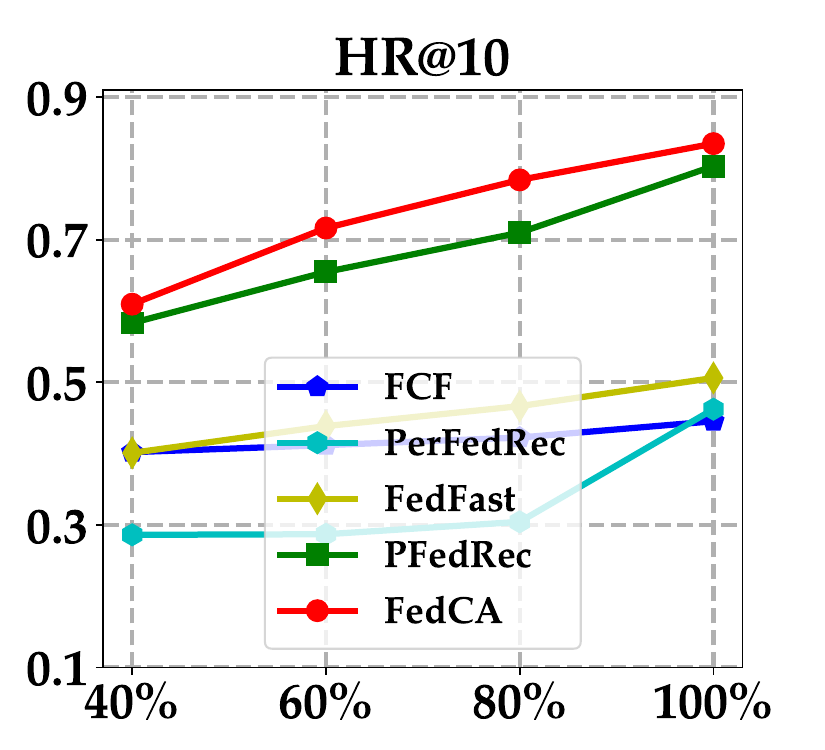}
    \caption*{(c) ML-1M}
   \end{subfigure}
        \begin{subfigure}{0.45\linewidth}
  \centering
    \includegraphics[width=\linewidth]{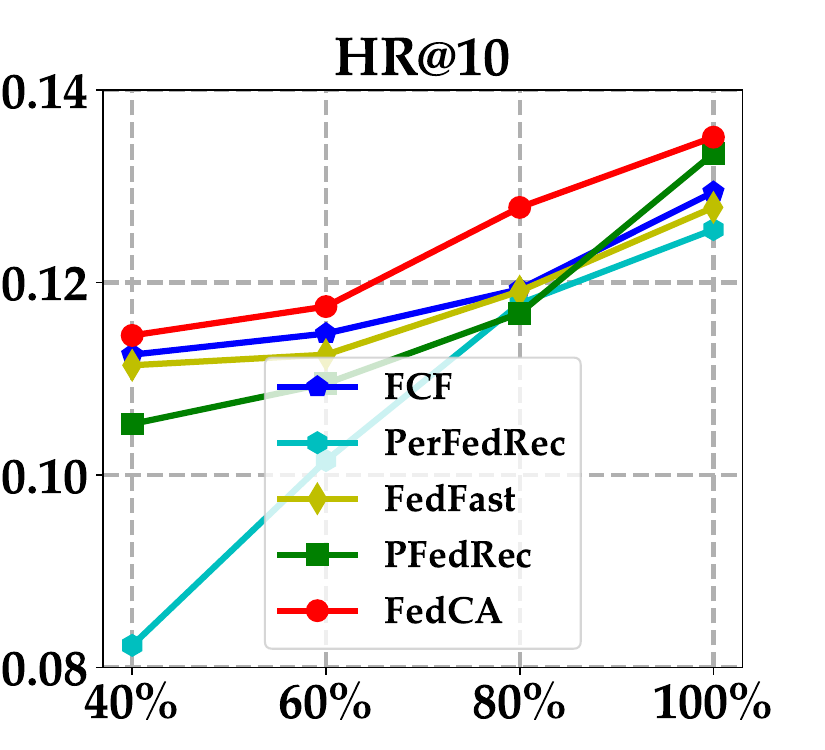}
    \caption*{(d) MC-100K}
   \end{subfigure}
    \caption{HR@10 results comparing our FedCA with FR baselines under varying train data ratios.}
    \label{fig:ratio}
\end{minipage}
\vspace{-0.6cm}
\end{wrapfigure}
\textbf{Overall Performance.} Table~\ref{tab:main_results} presents the results of our model compared to baselines using two backbones, evaluated in terms of HR@10 and NDCG@10 across four datasets. From the experimental results, we can observe that: (1) compared to local training, general FL methods (FedAvg, FedAtt and pFedGraph) and FR models (FCF, PerFedRec, FedFast, PFedRec) demonstrate better predictive performance, indicating the effectiveness of various aggregation mechanisms in federated settings. (2) by comparing different aggregation mechanisms, it can be noticed that both similarity-based aggregation (PerFedRec, FedAtt, and pFedGraph) and dissimilarity-based aggregation (FedFast) can achieve effective knowledge aggregation in federated settings. This suggests the motivation of our model to combine similarity and complementarity. (3) our method outperforms other baseline models, indicating that our FedCA, compared to solely using similarity for aggregation as borrowed from federated vision, is more suitable for aggregating embedding tables in FR tasks.

\begin{wrapfigure}{r}{0.5\textwidth}
\vspace{-0.4cm}
\begin{minipage}{\linewidth}
  \centering
  \begin{subfigure}{0.45\linewidth}
  \centering
    \includegraphics[width=\linewidth]{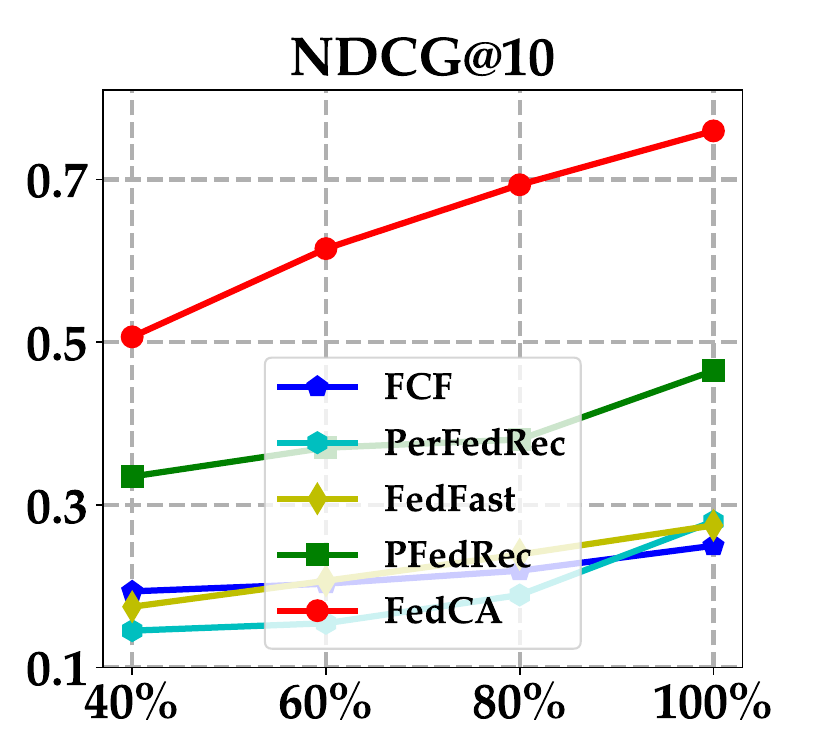}
    \caption*{(a) ML-100K}
   \end{subfigure}
     \begin{subfigure}{0.45\linewidth}
  \centering
    \includegraphics[width=\linewidth]{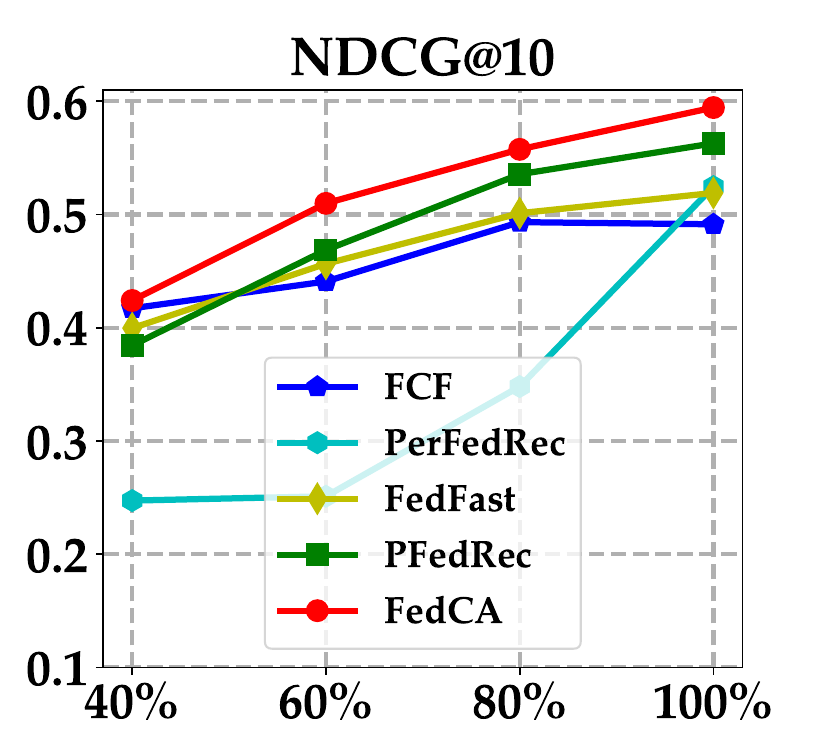}
    \caption*{(b) Filmtrust}
   \end{subfigure}
        \begin{subfigure}{0.45\linewidth}
  \centering
    \includegraphics[width=\linewidth]{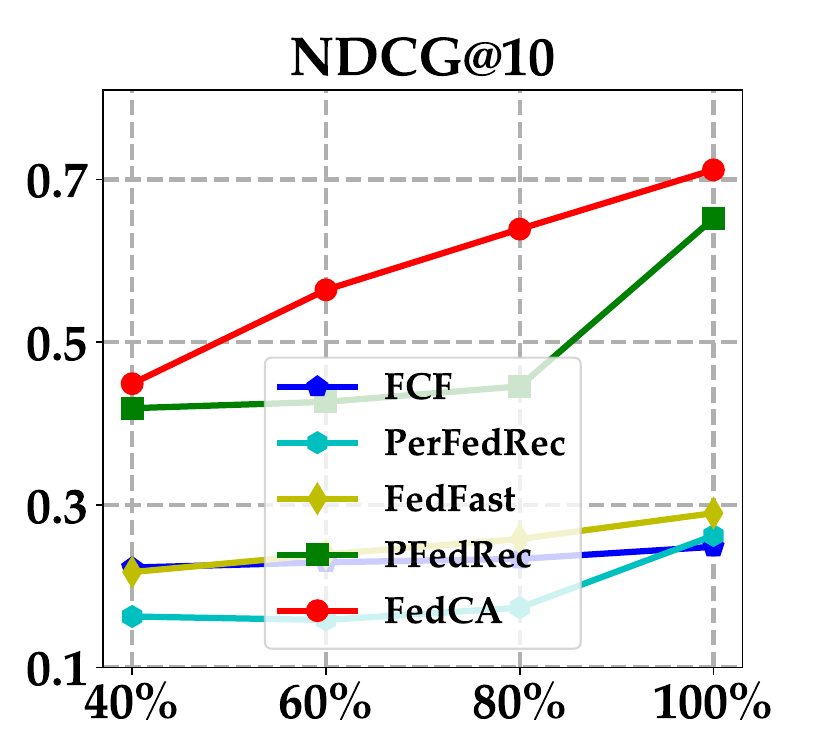}
    \caption*{(c) ML-1M}
   \end{subfigure}
        \begin{subfigure}{0.45\linewidth}
  \centering
    \includegraphics[width=\linewidth]{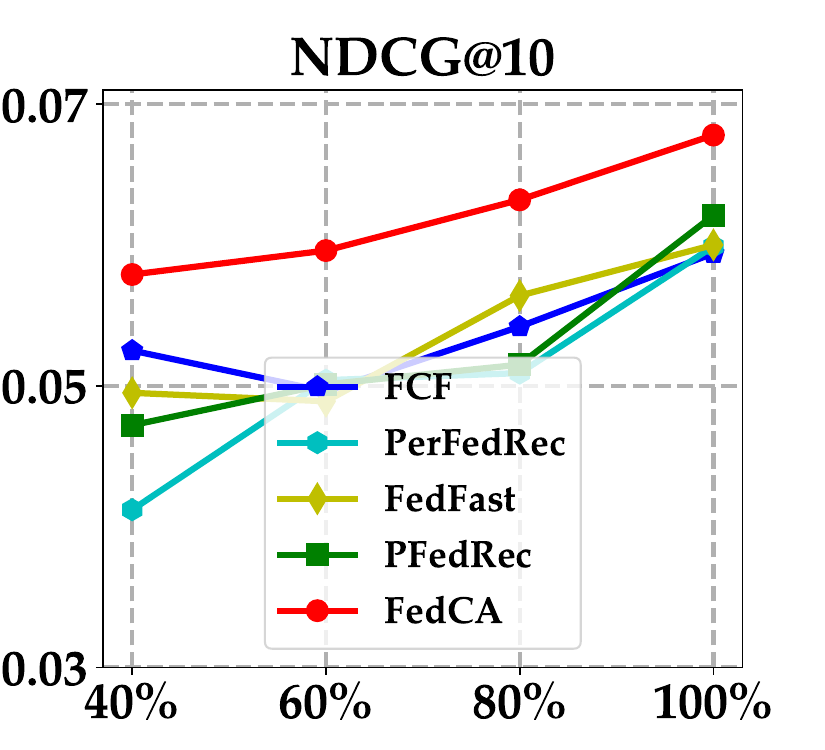}
    \caption*{(d) MC-100K}
   \end{subfigure}
    \caption{NDCG@10 results comparing FedCA with FR baselines under varying train data ratios.}
    \label{fig:ratio_ndcg}
\end{minipage}
\vspace{-0.4cm}
\end{wrapfigure}

\textbf{Robustness to varying sparsity of train data.} Recall from the empirical analysis in Section~\ref{sec:emp} that solely using similarity for aggregating embedding tables in FRs can lead to embedding skew issue. This means that as the aggregation process, those already trained embeddings improve while untrained ones remain random or degrade, ultimately failing to make predictions on future items. Our composite aggregation aims to alleviate this problem in FRs by combining similarity and complementarity to enhance untrained embeddings. Hence, theoretically, even with limited training data, our method can still achieve good generalization on test sets. To further explore the efficacy of our model, we evaluate the robustness of four FR methods instantiated on PMF backbone under different train data sparsity (40\%, 60\%, 80\% and 100\%). The experimental results on HR@10 and NDCG@10 are shown in Fig.~\ref{fig:ratio} and Fig.\ref{fig:ratio_ndcg}.

The results suggest that our method consistently outperforms other baselines under different levels of training data sparsity, directly demonstrating the effectiveness of our composite aggregation mechanism for aggregating embedding tables in FRs. Specifically, when the sparsity of train data is
at 40\%, our model significantly outperforms the baselines on the ML-100K and MC-100K datasets. This indicates that combining similarity and complementarity for aggregating embedding tables is highly effective for FR tasks, especially when training data is very limited.

Besides, we observed that the cluster-based aggregation method (PerFedRec) performs the worst under sparse data conditions (40\%) and exhibits very unstable learning process during the training iterations. This is primarily because existing clustering methods (such as K-Means~\cite{boutsidis2010random}) require careful selection of the number of clusters, as different datasets have varying client scales. Moreover, with limited data, it is challenging to accurately measure the similarity of each client, ultimately leading to the failure of cluster-based aggregation methods. This finding is consistent with the very recent work~\cite{co_cluster_2024}. In contrast, our method formulates the process into a unified optimization loss to smoothly select more similar clients, effectively achieving the benefits of cluster-based aggregation without the need for manual parameter tuning.

\begin{wrapfigure}{r}{0.6\textwidth}
\vspace{-0.5cm}
\begin{minipage}{\linewidth}
  \centering
  \begin{subfigure}{0.3\linewidth}
  \centering
    \includegraphics[width=\linewidth]{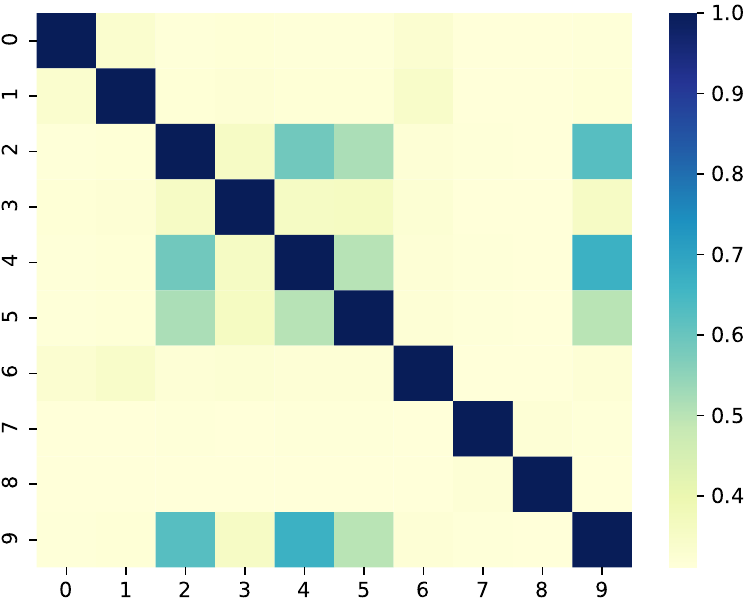}
    \caption*{(1) Similarity}
   \end{subfigure}
     \begin{subfigure}{0.3\linewidth}
  \centering
    \includegraphics[width=\linewidth]{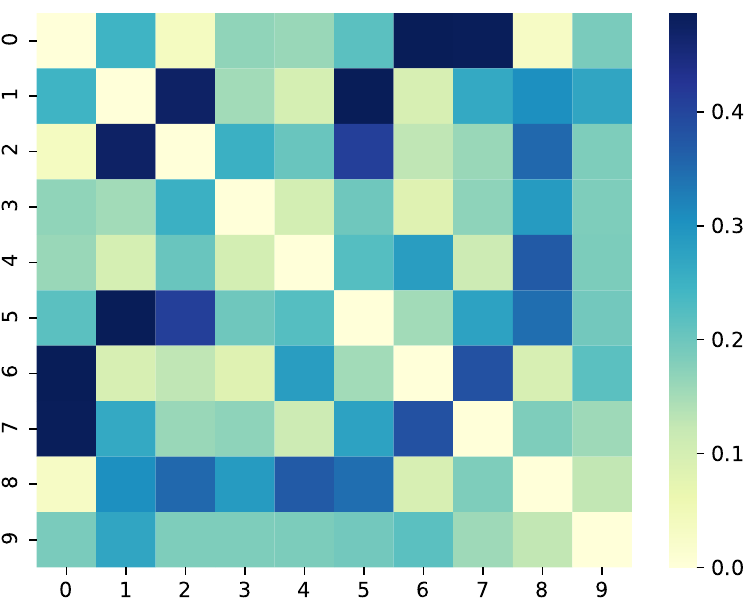}
    \caption*{(2) Complementary}
   \end{subfigure}
    \begin{subfigure}{0.3\linewidth}
  \centering
    \includegraphics[width=\linewidth]{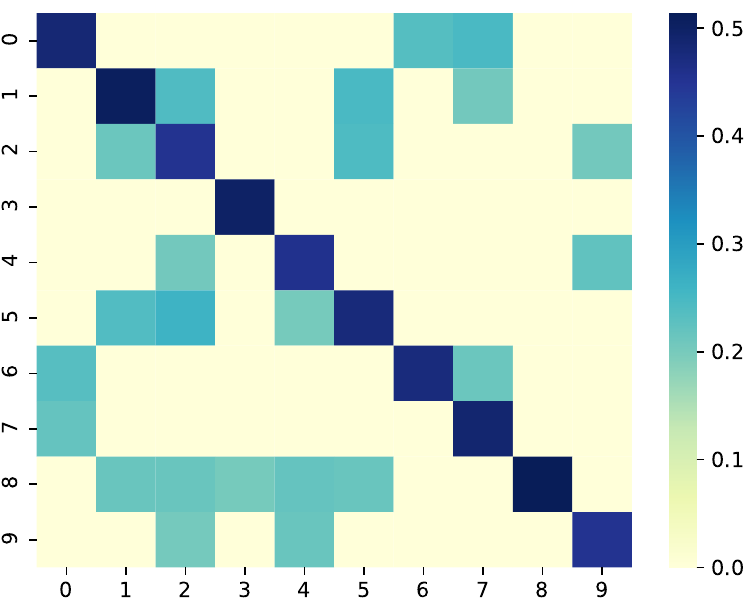}
    \caption*{(3) Composite}
   \end{subfigure}
   \vspace{-0.1cm}
    \caption*{(a) ML-100K}
  \begin{subfigure}{0.3\linewidth}
  \centering
    \includegraphics[width=\linewidth]{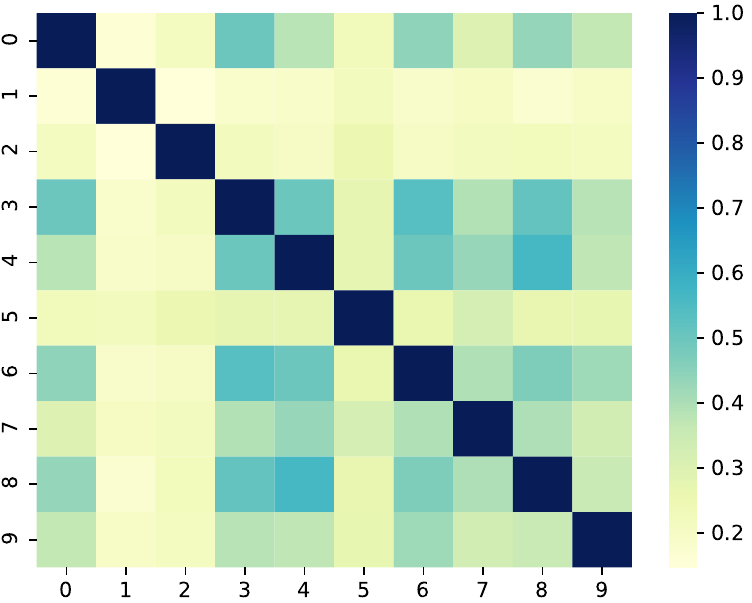}
    \caption*{(1) Similarity}
   \end{subfigure}
     \begin{subfigure}{0.3\linewidth}
  \centering
    \includegraphics[width=\linewidth]{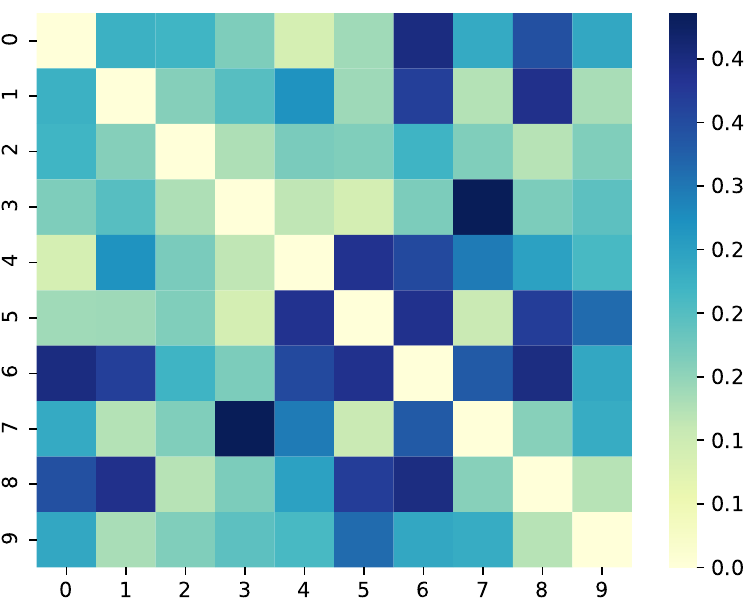}
    \caption*{(2) Complementary}
   \end{subfigure}
    \begin{subfigure}{0.3\linewidth}
  \centering
    \includegraphics[width=\linewidth]{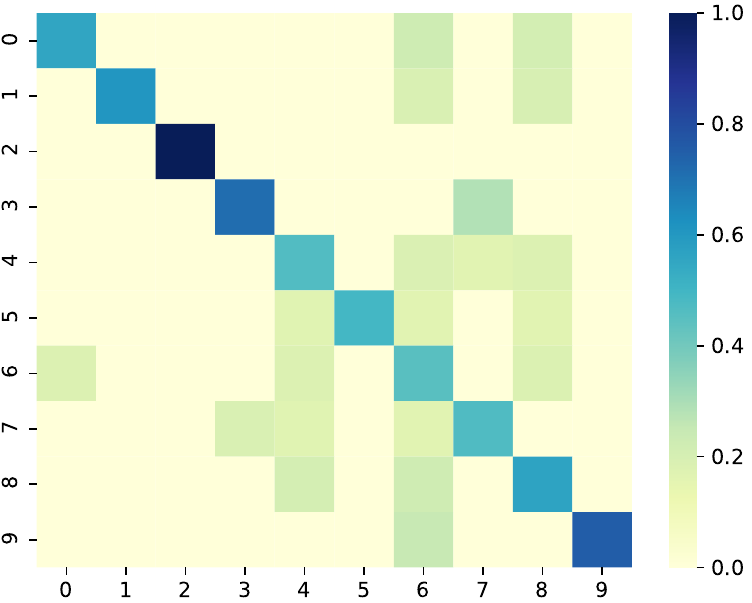}
    \caption*{(3) Composite}
   \end{subfigure}
\vspace{-0.1cm}
    \caption*{(b) Filmtrust}
    \caption{Visualization results regarding similarity, complementarity, and composite aggregation weights. }
    \label{fig:vis}
\end{minipage}
\vspace{-0.4cm}
\end{wrapfigure}

\textbf{Visualizing composite aggregation.} Fig.~\ref{fig:vis} presents the visualization results of the similarity matrix $\mathbf{S}$, the complementarity matrix $\mathbf{C}$, and the composite aggregation weights $\mathbf{W}$, reflecting their mutual influence in the overall loss function in Eq.(~\ref{eq:fedca}). We randomly select 10 clients for this demonstration. From the resutls, it can be observed that the composite aggregation weight effectively balances similarity and complementarity, accommodating both the model similarity and data complementarity among clients. It tends to favor clients with both high similarity and complementarity, where similarity ensures the consistency of the embedding distribution for interacted items among clients, and complementarity aims to enhance the embeddings of non-interacted items from other clients, which is similar to the classical user-based collaborative filtering notion~\cite{usercf}.

\begin{wrapfigure}{r}{0.55\textwidth}
\vspace{-0.3cm}
\centering
\captionsetup{type=table}
  \setlength\tabcolsep{1.2pt}
  \begin{tabular}{c|cc|cc}
    \toprule
      Proximal & \multicolumn{2}{c}{ML-100K}  & \multicolumn{2}{c}{ML-1M} 
      \\ \cline{2-5}
       term  & {HR@10}  & {NDCG@10}  & {HR@10}  & {NDCG@10}   \\
    \midrule
      \textit{w}  &  0.8469  & 0.7274  & 0.8168 & 0.6801   \\
      \textit{w/o} & \textbf{0.8738}  & \textbf{0.7597}  & \textbf{0.8348} & \textbf{0.7118}  \\
    \bottomrule
  \end{tabular}
    \caption{Results of the proximal term in FR tasks.}
  \label{tab:proximity}
  \vspace{-0.3cm}
\end{wrapfigure}

\textbf{Exploring the proximal term during local training.} We explored the effectiveness of the proximal term in FR tasks, which is widely used in federated vision domains. Table~\ref{tab:proximity} presents the experimental results with (w) and without (w/o) the proximal term based on the local task-specific loss in Eq.(~\ref{eq:bce}). It can be observed that not utilizing the proximal term constraint yields higher predictive performance in FR tasks. This indicates that recommendation tasks require stronger personalization at the local client level. Therefore, it is not necessary to enforce the local model to be as similar as to the global model. Instead, it should fully learn the personalized local model of each local client.

\textbf{Effect of the interpolation method during local inference.} We validated the efficacy of the proposed interpolation method in local inference. From the results in Table.~\ref{tab:ir}, it can be seen that the local model achieved optimal performance when $\rho = 0.8$ and $\rho = 0.9$ on the ML-100K and ML-1M datasets, respectively. This suggests that during local inference, a balance should be struck between the global aggregated model at current round and the local model at last round. This balance helps mitigate the spatial misalignment issue caused by the client-specific user embedding $\mathbf{p}_u$ in FR tasks, which is the main difference compared to federated vision domain.

\begin{wrapfigure}{r}{0.55\textwidth}
\vspace{-0.4cm}
\centering
\captionsetup{type=table}
  \setlength\tabcolsep{1.1pt}
  \begin{tabular}{l|cc|cc}
    \toprule
      \multirow{2}{*}{{$\rho$}} & \multicolumn{2}{c}{ML-100K}  & \multicolumn{2}{c}{ML-1M} 
      \\ \cline{2-5}
         & {HR@10}  & {NDCG@10}  & {HR@10}  & {NDCG@10}   \\
    \midrule
      0.5 & 0.6681  & 0.4930  & 0.6833 & 0.4892  \\
      0.6  &  0.7031  & 0.5295  & 0.7028 & 0.5158   \\
      0.7  & 0.7434  & 0.5675 & 0.8025 & 0.6711 \\ 
      0.8 & \textbf{0.8738}  & \textbf{0.7595}  & 0.8278  & 0.6949 \\
      0.9  &  0.8611  & 0.7432  & \textbf{0.8348} & \textbf{0.7118}  \\
     1.0  & 0.7922    & 0.6329  & 0.8008  & 0.6918 \\
    \bottomrule
  \end{tabular}
    \caption{Results for the interpolation method during local inference.}
  \label{tab:ir}
  \vspace{-0.4cm}
\end{wrapfigure}

\textbf{Ablation study.} To validate the effectiveness of the proposed composite aggregation mechanism, we decompose the overall optimization loss in Eq.~\ref{eq:fedca} into three components: client-specific task loss $\mathcal{L}_u$, model similarity loss $\mathcal{F}_s$, and data complementarity loss $\mathcal{F}_c$. Since the server cannot access local data, the client-specific loss is represented by the first squared loss term in Eq.~\ref{eq:optimize_w}. From the experimental results shown in Table~\ref{tab:ablation}, we can observe the following important conclusions.

\begin{wrapfigure}{r}{0.6\textwidth}
\vspace{-0.4cm}
\centering
\captionsetup{type=table}
  \setlength\tabcolsep{1.1pt}
  \begin{tabular}{l|cc|cc}
    \toprule
      \multirow{2}{*}{{Loss type}} & \multicolumn{2}{c}{ML-100K}  & \multicolumn{2}{c}{ML-1M} 
      \\ \cline{2-5}
          & {HR@10}  & {NDCG@10}  & {HR@10}  & {NDCG@10}   \\
    \midrule
      $\mathcal{L}_u$ & 0.4878  & 0.2786  & 0.4912 & 0.2751  \\
      $\mathcal{L}_u$ + $\mathcal{F}_c$  &  0.8431  & 0.6983  & 0.8049 & 0.6585   \\
      $\mathcal{L}_u$ + $\mathcal{F}_s$  &  0.8653  & 0.7457  & 0.8103 & 0.6987   \\
      $\mathcal{L}_u$ + $\mathcal{F}_c$ + $\mathcal{F}_s$  & \textbf{0.8738}  & \textbf{0.7597} & \textbf{0.8348} & \textbf{0.7118} \\ 
    \bottomrule
  \end{tabular}
    \caption{Results for using different loss combinations.}
  \label{tab:ablation}
\end{wrapfigure}

Optimizing the aggregation process using only the $\mathcal{L}_u$ component is analogous to the weighted aggregation in FedAvg. This variant helps demonstrate the basic aggregation without additional constraints. Based on that, considering complementarity $\mathcal{F}_c$ and similarity $\mathcal{F}_s$ modules separately both improve model performance, indicating that both factors play important roles in the aggregation process. Our proposed composite aggregation method utilizes a unified optimization framework to integrate similarity $\mathcal{F}_s$ and complementarity $\mathcal{F}_c$. This approach significantly enhances the effectiveness of aggregating embedding tables in FR tasks. 

\vspace{-0.2cm}

\section{Conclusion}

\vspace{-0.2cm}

This work first rethinks the fundamental differences between federated vision and FR tasks. Specifically, the federated vision community primarily utilizes structured parameters (e.g., convolutional neural networks) for federated optimization, whereas FR tasks mainly employ one-to-one item embedding tables for personalized recommendations. This key difference renders similarity-based aggregation borrowed from federated vision domain ineffective for aggregating embedding tables, leading to embedding skew issues. To address the above challenge, we introduce a composite aggregation mechanism tailored for FR tasks. Specifically, by combining model similarity and data complementarity within a unified optimization framework, our approach enhances the trained embeddings of items that a client has already interacted with and optimizes the non-trained embeddings of items the client has not interacted with. This enables effective prediction of future items. Besides, we explore the ineffectiveness of the proximal term on personalized preferences in FR tasks and propose an interpolation method to alleviate the spatial misalignment issue in FRs.

This research specifically proposes a promising composite aggregation framework for FR tasks. It is a model-agnostic, plug-and-play module that can be seamlessly integrated into mainstream FR models. However, we need to manually adjust the weight allocation for similarity and complementarity in this work. These limitations can be alleviated by using automated machine learning techniques~\cite{feurer2015efficient} to learn the weight allocation adaptively in future studies. Besides, exploring more suitable model similarity and data complementarity mechanisms for FR tasks is also a promising research direction.


\bibliographystyle{unsrt}
\bibliography{fedca}

\end{document}